"How lonely are you?" The role of social contacts and farm characteristics in farmers' self-reported feelings of loneliness—and why it matters

**Authors:**

Victoria Junquera*[1,2], Daniel I. Rubenstein[2], and Florian Knaus

[1] High Meadows Environmental Institute, Princeton University, Princeton, NJ, USA
[2] Department of Ecology and Evolutionary Biology, Princeton University, Princeton, NJ, USA
[2] Department of Environmental Systems Science, ETH Zurich, Zurich, Switzerland

* Victoria Junquera

Email: vjunquera@princeton.edu

**ORCID**

Victoria Junquera: 0000-0003-0402-3659
Daniel I. Rubenstein: 0000-0001-9049-5219
Florian Knaus: 0000-0003-3919-4730

**"How lonely are you?" The role of social contacts and farm characteristics in farmers' self-reported feelings of loneliness—and why it matters**

## ABSTRACT

Loneliness and social isolation among farmers are growing public health concerns. The contributing factors are manifold, and some of them are linked to structural change in agriculture, for instance because of higher workloads, rural depopulation, or reduced opportunities for collaboration. Our work explores the interconnections between loneliness, social contacts, and structural factors in agriculture based on a survey of 110 farm managers in the mountain region of Entlebuch, Switzerland combined with agricultural census data. We use path analysis, in which loneliness is the main outcome, and social contacts are an explanatory and explained variable. We find that 3% of respondents report that they feel lonely frequently or very frequently, and the rest sometimes (20%), rarely (40%) or never (38%). Managers with higher workloads report feeling lonely more frequently, and this relationship is direct, as well as indirect because of less frequent social contacts. However, physical isolation is not a significant predictor of loneliness. Moreover, short food supply chains correlate with less frequent loneliness feelings. Our study sheds light on the effects that structural change can have on the social fabric of rural areas.

## INTRODUCTION

Loneliness, social isolation, and mental health issues among farmers are growing public health concerns (Braun, 2019; Eisenreich and Pollari, 2021). A recent systematic review shows that 71% of articles report worse mental health in farmers than in the general population, compared to 18% that found the opposite and 11% that found no difference (Yazd et al., 2019). Studies also show significantly higher suicide rates among farmers compared to non-farmers (Kennedy et al., 2021). This topic is increasingly drawing the attention of mainstream media in Switzerland (Schneider Wermelinger, 2012; Stamm, 2022; SWI, 2018) and around the world (Williamson 2023; Morgan 2023; Szathmary 2023; Bhattacharya, Raj, and Pathak 2023).

Farming has occupational specificities that can be physically and emotionally taxing, and elevated levels of stress, physical injury, and suicide have been reported within the industry (Wheeler et al., 2023). High work demands and expectations coupled with low control over external stressors such as disease outbreaks, weather events, price fluctuations or changes in agricultural policies can lead to high levels of stress and mental health issues (Lunner Kolstrup et al. 2013). Farming is also a notoriously lonely occupation (Wheeler et al., 2023), and loneliness and social isolation are predictors of physical and mental health problems (Richard et al., 2017).

Structural change in agriculture—increasing farm size and intensity, decreasing farm numbers, growing specialization (Neuenfeldt et al., 2019), and gradual replacement of family with hired labor (Lobao and Meyer, 2001)—could be making matters worse. Agriculture in Europe and much of the rest of the world has undergone sustained structural change over the last decades, although with significant variability between and within countries (Giannakis and Bruggeman, 2015). In the European Union and Switzerland, the decline in farm numbers has disproportionately affected smaller farms, with the share of larger farms increasing (Möllers, Buchenrieder, and Csáki 2011; BfS 2023d), while the extent of agricultural area has remained approximately steady (BfS, 2023b; eurostat, 2023).

The process of sustained structural change could have negative social effects, for example if growing farm size and intensity result in longer work hours and less time for social exchange, or if declining rural populations translate into fewer opportunities for interaction (Junquera et al., 2022). The hypothesis that structural change negatively affects social capital was first formulated by Water Goldschmidt (1946), who found an inverse relationship between farm size and community participation. Since then, many studies have analyzed the relationship between changing farm structures and changing social capital in rural areas (Lobao and Meyer, 2001). Growing mechanization tends to reduce interactions between farmers in the field and the need for cooperation (Burton et al., 2006; Sutherland and Burton, 2011). Beyond the farm, streamlining and specialization across the supply chain may increase production efficiencies at the expense of traditional places of social exchange; an example is the replacement of  municipal milk collection points, where farmers met daily, with decentralized farm-gate milk collection (Junquera et al., 2022). Structural change can also intensify stress factors associated with managing large workloads, long working hours, time pressure, or juggling between on and off-farm work (Braun, 2019; Perceval et al., 2018).

There are a number of qualitative and quantitative studies examining the link between farmers' social connections and mental health issues and/or social isolation (Parent, n.d.; Perceval et al., 2018; Stain et al., 2008), or the association between farm and farmer characteristics and mental health (Henning-Smith et al., 2022; Lunner Kolstrup et al., 2013; Rudolphi et al., 2020; Yazd et al., 2019). However, few studies have quantitatively analyzed how farmers' feelings of loneliness, social connections, and farm structural factors relate to each other—the question that we examine in this work.

We focus on the Swiss mountain region of Entlebuch, which has a strong economic base around dairy and livestock farming, while showing strong signs of structural change over the last years and decades. A farm household survey conducted in 2018 asked about farmers' frequency of social contacts across 64 categories, ranging from friends, social associations, and social events, to administrative and professional contacts at local, regional, and national level, as well as the change in frequency of such contacts since the year 2000. It also asked farmers to rate how frequently they felt lonely.

An earlier study examined the link between (changing) farm structures and farmers' (changing) social connections (Junquera et al., 2022). Qualitative results showed that farmers felt they had more work than two decades ago, associated with a growing administrative burden and the necessity to produce more in order to stay economically afloat. This left less time and fewer opportunities for exchanges with friends and colleagues, which was exacerbated by the decreasing number of farms in the region. Quantitative results showed a reduction in the frequency of personal contacts (but an increase in professional and administrative contacts) over the last two decades and a correlation between higher workloads and fewer personal contacts.  These findings pointed to a net loss in social capital in the region.

The present work is based on the same dataset and analyzes as of yet unpublished data to examine the interconnection between feelings of loneliness, farm structures, social contacts, and other factors that have been suggested to influence loneliness, including physical isolation.

We analyze: 1) the distribution of self-reported frequency of loneliness feelings in the sampled population, 2) whether there are trends in loneliness across elevation zones and farm types, and 3) the relationship between loneliness, physical isolation, social contacts, and other farm and farm manager characteristics, such as farm size, intensity, or age.

To conduct our analysis, we use descriptive statistics and regression analysis. Given that social contacts are both an explanatory and an explained variable, we use path analysis, which is a form of multiple regression and a special case of structural equation modeling (Rosseel, 2012).

## CONCEPTUAL FRAMEWORK AND HYPOTHESES

### Dimensions of loneliness and contributing factors

Our theoretical framework is based on the definition and dimensions of loneliness proposed by Wheeler et al. (2023). Loneliness is a subjectively experienced, negative emotion related to a person's perception that their social network is deficient either qualitatively or quantitatively (Wheeler et al., 2023). Wheeler et al. (2023) define three types of loneliness—social, emotional, and cultural—which are somewhat overlapping and mutually reinforcing.

*Social* loneliness is a perceived deficit in the quality and quantity of social connections and is most closely associated with social isolation. Social isolation might reflect absence of a wide or close enough social circle (number and type of contacts), but also a lack of *opportunity* to engage with or expand that circle (frequency of contacts), e.g., due to lone and long working hours, physical separation, or restricted social networks (Wheeler et al., 2023). Loneliness and social isolation are distinct phenomena: loneliness is a subjective emotion, whereas social isolation is an objective measure and while they are closely connected, each can exist in the absence of the other (Wheeler et al. 2023). Social loneliness among farmers can be associated with long working hours and the resulting lack of free time for leisure and social activities, working alone, or geographical isolation from peers and friends, resulting in an inability to maintain existing social relationships or build new ones (Wheeler et al., 2023).

*Emotional* loneliness refers to lacking intimate relationships and close contacts an individual can confide in. An individual may have a large social circle but still experience emotional loneliness if relationships are deemed to be lacking in quality. Emotional loneliness may also be caused by family tensions, relationship difficulties, etc. (Wheeler et al., 2023).

*Cultural* loneliness arises from the sense that the cultural group within which an individual identifies  is poorly understood or marginalized by mainstream society. It arises from a sense of disconnection, perceived lack of public understanding, or public pressure or criticism against farmers (Wheeler et al., 2023). Cultural loneliness may be addressed by improving connections and understandings with different social groups, such as the wider non-farming public (Wheeler et al., 2023).

Aside from social contacts, there is evidence that *stress* has a direct, two-way, and mutually-reinforcing causal relationship with loneliness: high stress is a co-cause of loneliness (Campagne, 2019) and social isolation increases stress levels (Jones-Bitton et al., 2020). Certain elements specific to a farming life and culture contribute to high levels of stress among farmers. These include a complex and physically demanding work environment, long work days, growing reliance on technology that may decrease human-nature contact and run counter to traditional farming practices (Lunner Kolstrup et al., 2013), as well as ideas around stoicism, strength, self-reliance, "hard work", traditional gender roles, or family expectations around farm inheritance, succession, and continuity (Alston and Kent, 2008; Wheeler et al., 2023). Additional stress factors are high levels of debt and administrative burden (Ritzel et al., 2020) and high exposure to external stressors, such as economic and climate unpredictability, animal disease, or changes in government regulations (Jones-Bitton et al., 2020; Lunner Kolstrup et al., 2013).

**Relationship between farm structures and social contacts**

An earlier study examined the connection between farmers' social contacts and farm/er characteristics in the study area. Controlling for farm size, farm managers with higher workloads had fewer close social contacts (with family, friends, and colleagues), presumably because under time constraints, such "nonessential" contacts are most easy to give up (Junquera et al., 2022). However, controlling for workload, managers of larger and more intensive farms had a higher frequency of social contacts, possibly linked to higher social capital associated with the "good farmer" image (Burton, 2004; Sutherland and Burton, 2011). Moreover, farms with the local origin label ("*Echt Entlebuch*") reported more personal and professional local contacts, and organic farmers—a minority in Switzerland and especially in Entlebuch—were less socially connected locally but more connected regionally. This "otherness" effect may have also played a role in the lower than average close social contacts reported by managers of goat and sheep farms, which do not fit into the cattle tradition of the region. Among other variables analyzed, farmers' age and education did not exhibit significant correlation with social contact frequency, and neither did the elevation zone of the farm.

## Dimensions of loneliness and relationships captured by our dataset

The farm household survey asked farmers 'How often do you feel lonely?', without distinguishing between different aspects of loneliness. Hence, the answers are likely to reflect all (three) dimensions of loneliness—social, emotional, and cultural (Figure 1). We expect that our data explain *social* loneliness best, based on variables such as the frequency, type, and physical proximity of social contacts. Some aspects of *emotional* loneliness may also be captured by our data, for example in the frequency of social contacts with family and friends, although we did not ask specifically about the emotional closeness of social contacts. Conversely, we believe that our data are least able to explain *cultural* loneliness, which is a more complex feeling than social or emotional loneliness, although some farm characteristics, such as certifications or farm size and intensity, may have a bearing on emotional loneliness (see 2.4 Hypotheses). Beyond their direct correlation with social contacts, certain farm characteristics such as farm size, intensity, and workload could affect farmers' stress levels and thus could have an indirect impact on loneliness.

## Hypotheses

Based on the above insights on the dimensions of loneliness and on our available data, we assess whether loneliness is correlated with social contacts and/or with certain farm and farmer characteristics. We hypothesize that (Figure 1):

1. (A) A higher frequency and diversity of social contacts correlates negatively with loneliness by reducing farmers' social loneliness. (B) A higher frequency of *close* social contacts correlates negatively with loneliness by reducing farmers' emotional loneliness .

2. Workload is positively correlated with loneliness, (A) indirectly through its negative correlation with the frequency of close social contacts (Junquera et al., 2022), and (B) directly, e.g. because of its presumed association with higher stress levels.

3. A larger farm size and intensity correlate positively with social contact frequency (Junquera et al., 2022)—possibly due to the "good farmer" effect and increased social reputation (Burton, 2004)—and thus correlate negatively with loneliness, both social (because of more frequent social contacts) and cultural (because of higher acceptance by their peers and society at large).

4. Certifications and labels correlate negatively with loneliness (A) indirectly, because they are associated with more frequent social contacts, such as meetings with peers, local retailers, association members, and certifiers (Junquera et al., 2022), thus reducing social loneliness and (B) directly, because they may result in a higher sense of connectedness and belonging, thus reducing emotional loneliness (McDaniel et al., 2021)

5. A) Physical isolation, for example as measured by greater distance to neighbors or farm location at higher elevations, correlates with higher (social) loneliness because geographical isolation may reduce opportunities for social interaction. B) Moreover, more geographically proximate neighbors have a stronger negative impact on loneliness than more distant neighbors.

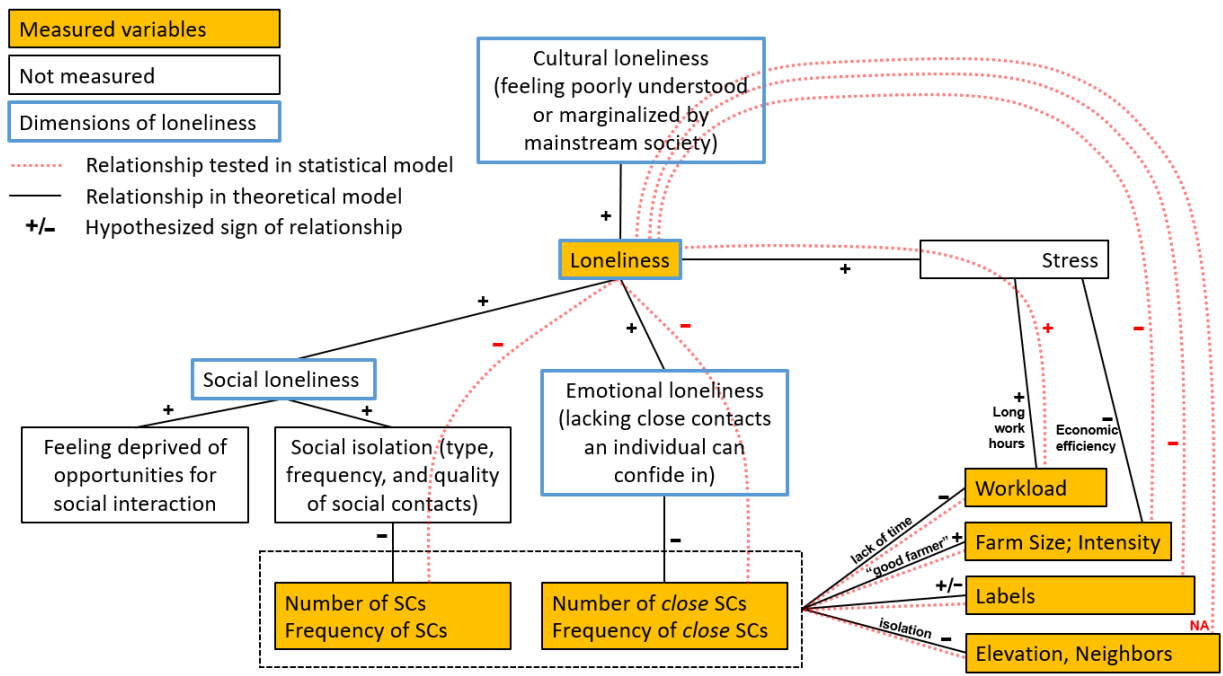

Figure 1: Diagram representing the dimensions of loneliness potentially captured by our measured variables, hypothesized causal relationships (black), and the implementation in the Path Analysis model (red dashes). Signs (+/-) represent hypothesized direction of correlation. SC: social contacts.

## METHODS

### Study area

The region Entlebuch is located in the Canton of Lucerne in central Switzerland and comprises the municipalities Entlebuch, Doppleschwand, Romoos, Schüpfheim, Escholzmatt-Marbach, Flühli, and Hasle (Figure2). It was designated as a UNESCO Biosphere Reserve in 2001 for its valuable natural and cultural landscapes (UNESCO, 2018). The UNESCO Biopshere Reserve Entlebuch (UBE) covers around 40,000 ha (roughly 10,000 acres) and is a mountain area with elevations ranging from 800 m.a.s.l. (mountain zone 1) to 2400 m.a.s.l. (mountain zone 4). Agricultural activities have a central cultural and economic role and comprise 21% of full-time jobs (LUSTAT, 2023) compared to the 2.3% Swiss average (BfS, 2023c). The centrality of agriculture is also reflected in the number of farms per 1000 inhabitants, which is 43 in Entlebuch, among the highest in Switzerland and well above Swiss-wide national average of 6 (Aargauer Zeitung, 2021). Most farms (98%) in Entlebuch are family owned and operated. Farmland consists almost entirely of pastures and meadows, as well as areas set aside for ecological compensation. Agricultural production centers around animal husbandry of dairy cows, goats, sheep, pigs, and veal mast. The region is unsuitable for cereal or horticultural production due to its rugged terrain and climate and soil characteristics. In 2017, 8% of UBE farms were certified organic (UBE, 2020), which is lower than the Swiss average of 15% (BfS, 2023d). The region also has its own local origin label (*Echt Entlebuch*).

The region Entlebuch has undergone strong structural changes in agriculture over the last decades. Between 2000 and 2017, farm numbers decreased by 22% to 986, while agricultural area in the region remained approximately stable around 1630 ha, so that average farm size increased by 28%, from 12.9 ha to 16.5 ha (BfS, 2023b).

Cantonal statistics show that the two smallest farm size classes (0-4.9 ha and 5-9.9 ha) have proportionally suffered the largest losses, and among these the smallest category was most affected; conversely, the fraction of the largest farm size classes (10-19.9 ha, and 20+ ha) has increased (Statistik Luzern, 2019, p. 118). In other words, smaller farms are disappearing at a faster rate than larger farms.

The drop in farm numbers was accompanied by a similar decline in the number of persons employed in agriculture (-21%), despite the fact that the agricultural area has remained the same (BfS, 2023b), with the highest decline among those working full time or near full time (-39%), meaning that the proportion of people working part time is increasing.

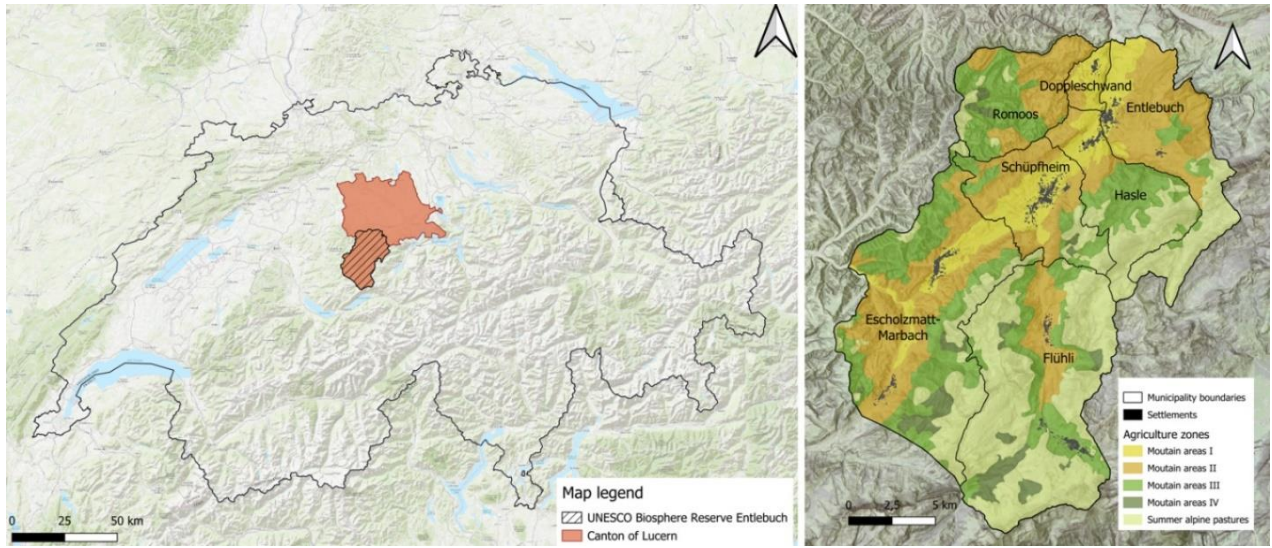

*Figure 2: Municipalities of the UNESCO Biosphere Reserve Entlebuch (UBE) (right) and location of study area in Switzerland (left). Data source: swisstopo (2023).*

## Data collection

We mailed paper-based surveys (7 pages long; in German) in November 2017 to 400 farms that were randomly sampled from all farms in the UBE *excluding* non-commercial farms, seasonal-only activities (such as alps), cooperatives, slaughterhouses, and excluding also 241 farms selected to receive a survey unrelated to this study around the same dates. The resulting sampling frame was N=643. Out of 400 surveys, we received N=110 usable responses corresponding to a response rate of 26%. Average farm size and distribution across mountain zones did not significantly differ from the UBE average (Junquera et al., 2022).

The survey was addressed to the farm manager or joint managers (e.g., couples, family members, etc.). It asked about the characteristics of the farm (type, certifications and labels, people living and working in the farm, employees and apprentices) and farm manager (age, formal and continued education), as well as the farm managers' current (in 2017) frequency of social contacts (annual, monthly, weekly, or daily) across 64 social contact groups (Supplementary Materials S1, Table S1-2), changes in the frequency of social contacts in each social contact group (increase, decrease, or no change) during the period 2000-2017, and the reasons for such changes, which were asked in a semi-structured question. The survey also asked 'How often do you feel lonely?' with possible answers on a 6-point Likert scale (never, rarely, sometimes, frequently, very frequently, always).

We also obtained farm-level agricultural census data for 2012 and 2017 for all farms in the Canton of Lucerne and matched our surveyed farms with those in the agricultural census by farm ID. Agricultural census data included address, mountain zone, total agricultural area (LN), standardized workload (SAK), and standardized livestock units (GVE) for each animal type. Based on GVE data, we assigned farm type based on the Swiss farming typology (Hoop and Schmid 201). A detailed description of the case study area and data collection is included in (Junquera et al. 2023).

In this work, we additionally calculated farm coordinates based on the physical address of farms and using a geocoder spreadsheet that translates address information into the Swiss coordinate system (swisstopo, 2023). We included all farms located in UBE and in surrounding municipalities within the Canton of Luzern (Schwarzenberg, Ruswil, Menznau, Malters, and Hergiswil bei Willisau). These municipalities surround the northern portion of the UBE, but the larger southern portion of the UBE borders with a different canton, for which we did not have agricultural census data. Among the N=110 farms in our sample, two had an inconclusive address and could not be geolocated.

## Analysis

### Variables

A detailed description of each variable used in the analysis is found in S1 (Table S1-1). For each respondent (N=110), self-reported frequency of loneliness feelings is reflected in the variable *lonely*, which is ordinal and has 6 categories (0=never, 1=rarely, 2=sometimes, 3=frequently, 4=very frequently, 5=always). A simplified version of this variable, *lonely_4cat* has only 4 categories by merging the three highest categories (0=never, 1=rarely, 2=sometimes, 3=frequently, very frequently, or always).

For each respondent, we calculate the total frequency of social contacts in 2017 within each of the 64 social contact groups (Table S1-2 in S1) (e.g., friends inside the UBE, local dancing organizations, cantonal agricultural inspection office, etc.) using weighting factors (1=annual, 12=monthly, 52=weekly, 365=daily). We also calculate change in social contact frequency between 2000-2017 within each social contact type (1=increase, 0=no change, -1=decrease).

We aggregate social contacts into 12 aggregate social contact categories, which we classify as personal or professional (Table S1-2). We also calculate *total* contacts (added across all 64 social contact groups). The variable *count* reflects the number of different social contacts (among the 64 groups). Because the social contact groups in our questionnaire are predominantly professional[1], the variable *count* is biased towards professional contacts, including administrative. For each aggregated social contact category, we calculate average contact frequency in 2017 and average change in social contacts between 2000–2017.

We categorize farm type based on livestock composition following the Swiss farm typology (Hoop and Schmid 201). We use binary variables to indicate whether the farm is certified organic (*BIO*) or sells its products under the local origin label *Echt Entlebuch* (*labEE*), or one of the two (*BIO_or_labEE*).

Based on farm coordinates, we calculate two proxies of physical isolation (number of neighbors within a given radius and distance to the nearest neighbor), which provide information about the separation between a farm in our sample from other surrounding farms. The first measure, *knn[X]dist*, reflects the distance X in kilometers between the farm and its Xth nearest neighbor, which we calculate for the $1^{st}$, $5^{th}$, and $10^{th}$ nearest neighbor. For example, knn1dist is the distance to the (first) nearest neighbor; similarly we calculate knn5dist and knn10dist. The second measure, *nn[X]m*, reflects the number of neighbors within a radius of X meters (500m, 750m, and 2000m).

*Methodological approach*

We use descriptive statistics to visualize the distribution of the frequency of loneliness feelings in the study population, as well as by elevation zone, farm type, and certifications (Section 4.1). We then plot the four-category loneliness variable (*lonely_4cat*) against various variables relating to social contacts, neighbors, farm size and intensity, and certifications (Section 4.2). In each of these figures, we examine whether any emerging trend is statistically significant by conducting a linear regression analysis between *lonely_4cat* and each variable.

In Section 4.3 we use Path Analysis (PA), a type of structural equation modeling (SEM) and a general case of multivariate regression where all endogenous variables are allowed to explain other endogenous variables (Rosseel, 2012). Here, we hypothesize that social contacts and other farm characteristics are predictors for, i.e., explain, loneliness. At the same time, social contacts are explained by farm characteristics such as workload, intensity, farm size, and certifications and labels (Junquera et al. 2022).

As model fit statistics for the PA, we use the model Chi-square, the Root Mean Square Error of Approximation (RMSEA), and the Comparative Fit Index (CFI) (Rosseel 2012). The closer the CFI is to 1, the better the fit of the model. A relative Chi-square (divided by the degrees of freedom) greater than 2 indicates poor fit (1 is perfect fit). A RMSEA lower than 0.05 indicates good model fit.

---

[1] Excluding political parties, 37 categories are professional, including administrative contacts, and 21 are personal; see S1.

We also conduct a multiple linear regression analysis, based on Ordinary Least Squares (OLS), using the same model in the PA but without the endogenous component—that is, with loneliness as the dependent variable and social contacts and farm characteristics as independent variables. The multiple linear regression allows us to compute the adjusted R-squared, which is a simple and broadly used measure of fit.

All variables in the regression analyses are centered (by subtracting the sample mean) and scaled (by dividing by the sample standard deviation) (Schielzeth, 2010).

We use the R statistical language and environment (R Core Team, 2023) for all data analyses. We use the *lm* function in base R for multiple regression analysis, and the *lavaan* package to conduct the PA (Rosseel, 2012). Estimated coefficients in *lm* are based on Ordinary Least Squares; estimated coefficients in *lavaan* are based on Maximum Likelihood and the NLMINB optimization method. The R code and data used for generating all results figures are included in the Supplementary Materials.

## RESULTS

### Self-reported frequency of loneliness in sampled population, by elevation and farm type

Out of 110 respondents, six did not answer the question about the frequency of loneliness. No farmer reports being always lonely (Figure 3A). Among the remaining 104, 38% (N=39) report that they never feel lonely and 40% (N=42) report that they rarely feel lonely, with 19% (N=20) reporting that they sometimes feel lonely. Only three farmers report frequent (N=2) or very frequent (N=1) feelings of loneliness. These three farms include a dairy milk farm in elevation zone 2, a veal mast farm in zone 3, and a goat/sheep farm in zone 3.

Managers of farms located at higher elevations report more frequent feelings of loneliness, but this trend is not statistically significant at p=0.1, nor are there significant differences in loneliness between farm types (Table S2-1).

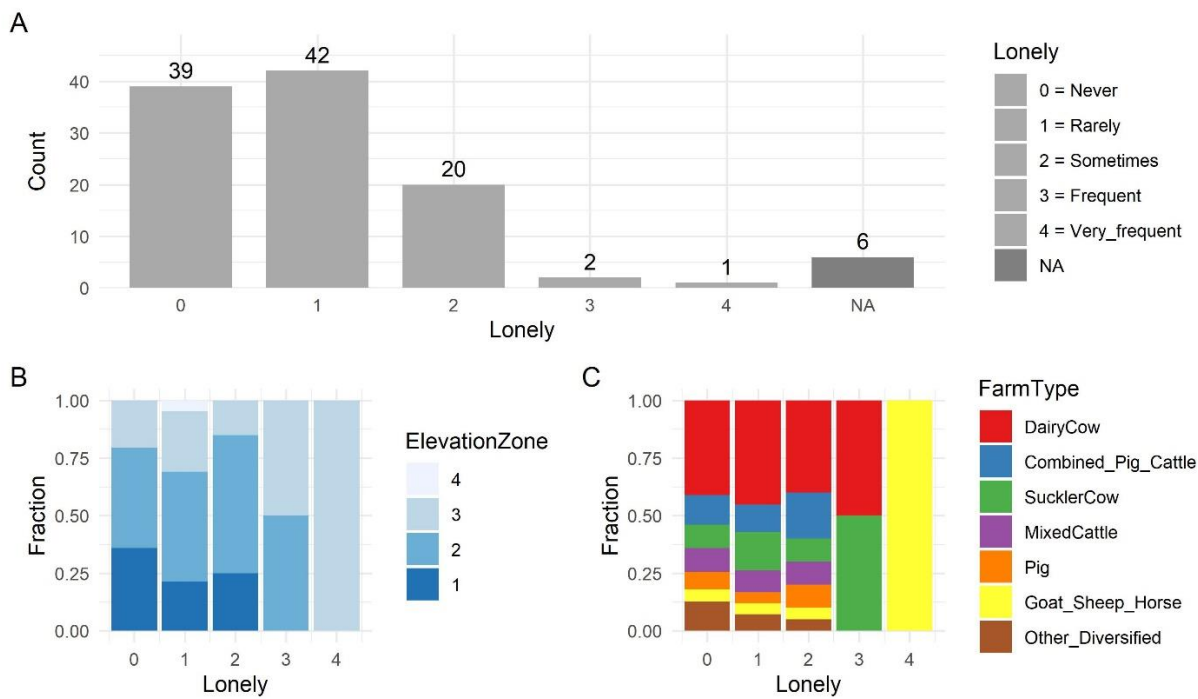

*Figure 3: Respondents' answers to the question "How often do you feel lonely?" (0=never, 1=rarely, 2=sometimes, 3=frequent, 4=very frequent, 5=always). (A) All respondents (N=110); (B) by farm elevation zone; (C) by farm type.*

## Descriptive statistics

We present descriptive statistics showing loneliness frequency against social contacts, farm characteristics, and their changes. Social contacts and farm characteristics refer to the year 2017. Changes in social contacts refer to the timeframe 2000–2017, whereas changes in farm size and intensity refer to the timeframe 2012–2017. Trendlines (shown only in Fig 4A, Total) and asterisks indicate whether OLS regression coefficients between loneliness and various regressors are significant at p<0.1.

### Total and disaggregated social contacts

Average frequency of loneliness decreases with increasing total social contacts, and this correlation is statistically significant (Figure 4A). A similar trend is also seen between loneliness and certain disaggregated social contact variables, notably contacts with colleagues and social organizations. Conversely, more frequent contacts with political parties correlate with more frequent loneliness.

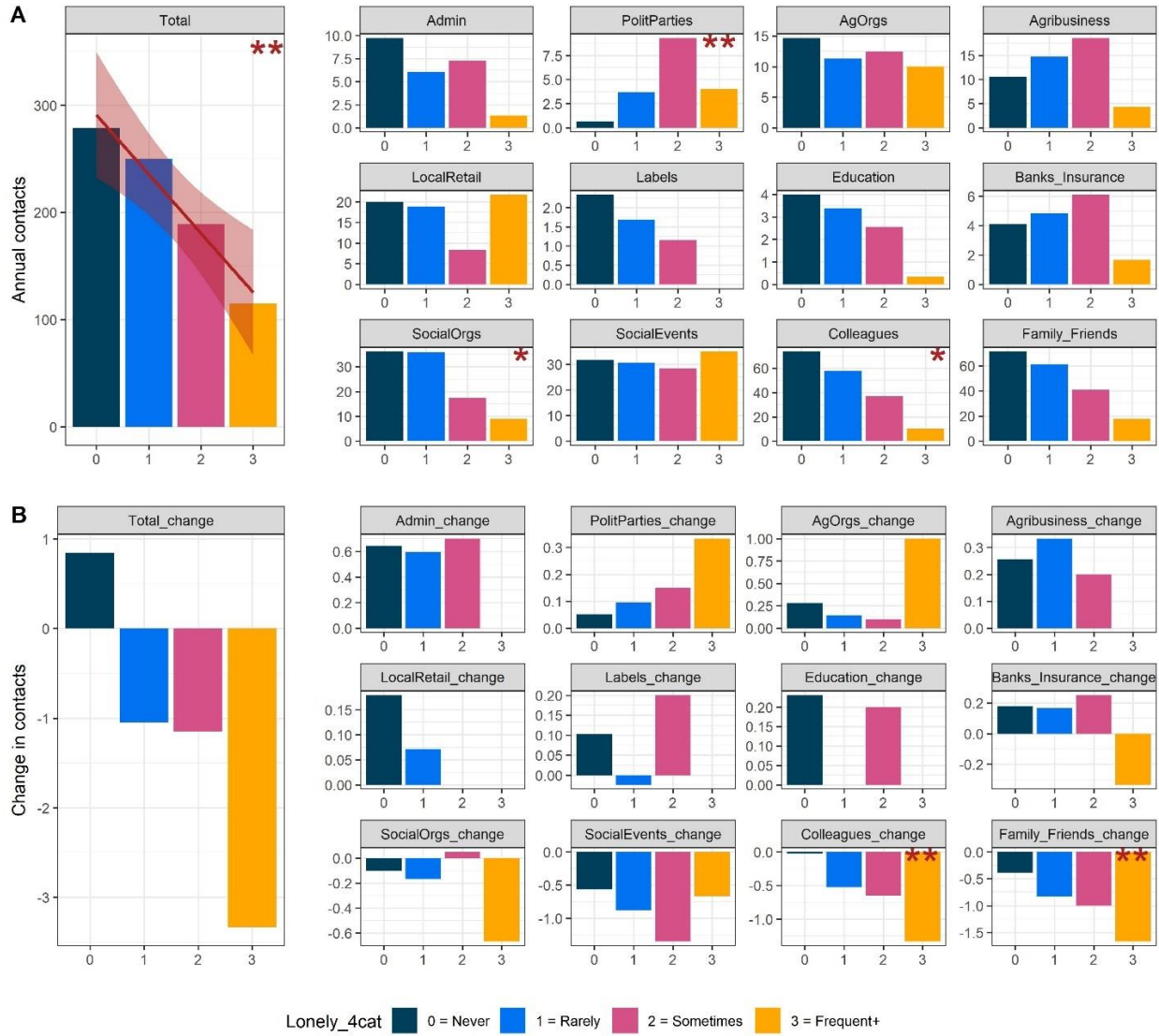

*Figure 4: (A) Loneliness (0=never, 1=rarely, 2=sometimes, 3=frequent or very frequent) plotted against (A) Annual frequency of social contacts in 2017, and (B) Change in frequency of social contacts since 2000. The line and confidence interval (shown only once) represent simple linear regression results; asterisks represent p-value of regression coefficient (' ' not significant at p<0.1, '\*' significant at  p<0.1, '\*\*' significant at p<0.05).*

## Changes in total social contacts

Farmers who report never to be lonely in 2017 had on average more frequent social contacts in 2017 than in the year 2000—that is, they had a net gain in social contact frequency (Figure 4B)—and farmers who report larger losses in social contacts also experienced more frequent loneliness, but this trend is not statistically significant. Among the disaggregated social contact categories, more frequent loneliness feelings are significantly correlated with higher losses in contacts with family, friends, and colleagues.

*Farm size and intensity*

Managers of larger farms as measured by workload (SAK) report higher loneliness levels, and this relationship is statistically significant ($p < 0.05$). The other farm size variables livestock units (GVE) and agricultural area (LN) are not significantly correlated with loneliness, and neither is intensity as measured by GVE/LN (Figure S3-1A). There is no significant association between the frequency of loneliness and changes in farm size and intensity (Figure S3-1B).

We examined the size and intensity, and their changes, in the three farms where managers reported the most frequent feelings of loneliness (see Section 4.1). The goat/sheep farm (where the manager reported being lonely very often) underwent a large expansion between 2012 and 2017 and roughly doubled in size (LN). Agricultural area in the suckler cow farm was reduced by about 25% in that period. All three farms had above-average increases in intensity (GVE/LN), and in the case of the dairy cow and goat/sheep farms this increase is more than one standard deviation above the average (not shown). This explains why the highest reported frequency of loneliness (3) corresponds to farms with large increases in GVE and GVE/LN in Figure S3-1B.

*Physical isolation*

The average distance to the nearest neighboring farm (*knn1dist*) is 0.16 km. We hypothesized that increasing distance from the nearest neighbor would result in higher frequency of loneliness. The trend exhibited by the variable *knn1dist* corresponds with this hypothesis, but the relationship is not significant at $p = 0.1$ (Figure S3-2A). We further hypothesized that the distance to the nearest neighbor has a larger impact on loneliness than the distance of subsequent neighbors. This is supported by Figure 7A, which shows no trend or correlation between loneliness and knn10dist or knn5dist. Farmers with more neighbors within a 500-meter radius are more likely to report that they are never lonely, but this trend cannot be seen for 2000 m or 750 m and it is also not significant for any radius. Overall, our results show that physical isolation of the farm as measured by the distance to the nearest neighbor or the number of neighbors within a certain radius is not a statistically significant predictor of loneliness.

*Labels and certifications*

Organic certification (BIO) exhibits no significant correlation or trend with reported loneliness (Figure S3-2B). There is, however, a growing proportion of farms with the local *Echt Entlebuch* label among farm managers reporting lower loneliness levels, but this trend is not statistically significant. When taking together organic certification or labels (*BIO_or_labEE*), the trend is significant.

## Path analysis

Model 1 of the PA examines the relationships between the frequency of loneliness and social contacts (personal, professional, and number of different contacts), farm structural variables (SAK, LN, GVE; or SAK, LN, intensity in Model 1B), certifications (*BIO_or_labEE*), and physical isolation (elevation zone; or number of neighbors in Model 1C). In a second model (Model 2) we add other farm characteristics such as employees, the number of family members working in the farm, and farmer characteristics including age, education, courses and advice. In all models, the variables personal and professional contacts are both explanatory and explained variables. After removing missing data, the number of observations is N=100 in Model 1 and N=94 in Model 2.

Results of Model 1 show that the frequency of both personal and professional contacts correlates negatively with the frequency of loneliness feelings, but the number of different contacts (*Count*) has a positive correlation with loneliness, all significant at p<0.1 (Figure 5; Table S2-2).

Workload (SAK) has a positive and significant correlation with loneliness. Conversely, smaller agricultural area (LN) correlates significantly with higher loneliness. However, intensity (GVE/LN) is not significantly correlated with loneliness (Model 1B, Table S2-3).

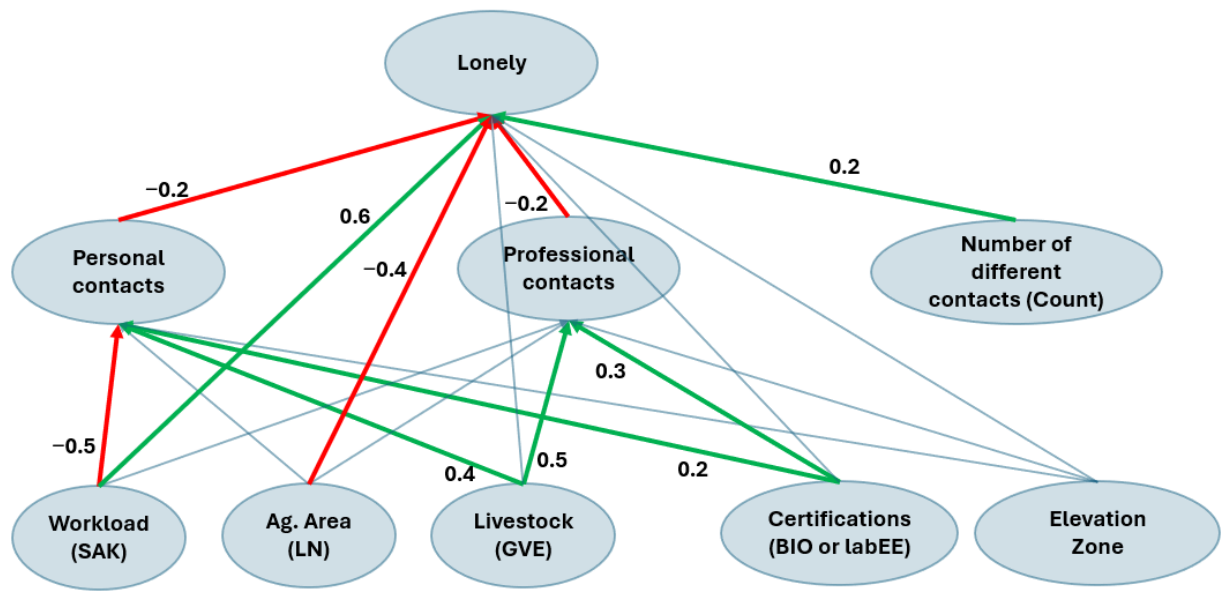

*Figure 5: Path Analysis Model 1. Numbers correspond to parameter estimates. Thick lines indicate statistically significant relationships at p<0.1, color indicates the sign of parameter estimates (red=negative, green=positive).*

The frequency of personal contacts is negatively correlated with workload (i.e., higher workloads correlate with fewer personal contacts), and positively correlated with livestock numbers and certifications. Professional contacts are not significantly correlated with workload, but they are significantly and positively correlated with livestock numbers and certifications.

Elevation zone is not significantly correlated with loneliness or with personal and professional contacts. Similarly, the distance to the nearest neighboring farm (Model 1C, Table S2-4) and other variables related to the number and distance of neighbors are also not significant (not shown).

Model 2 results (Table S2-5) show that none of the additional variables (employees, family members working in the farm, and managers' age, education, courses, and advice) is correlated with loneliness frequency. However, higher age is correlated with a lower frequency of personal contacts (coeff=−0.2, pval<0.05).

The multiple linear regression results (Tables S2-6&7) yield similar coefficients as the PA regression. The adjusted R-squared of the multiple regression is 10% for Model 1 and 9% for Model 2.

## DISCUSSION

The results of our study on loneliness in a Swiss faming population do not paint an overly alarming picture at first.  Out of the 104 farmers who answered the question "How often do you feel lonely?", only three respondents answered "frequently" (2%) or "very frequently" (1%). Among the rest, 19% answered "sometimes", and a large majority (78%) of respondents answered "never" (38%) or "rarely" (40%).

The three farmers who reported feeling lonely frequently or very frequently may seem a negligible number. Yet we consider them to be important outliers, who might be most vulnerable to mental health issues. A closer inspection revealed that these farmers have considerably increased their farms' intensity over the last five years, suggesting perhaps an attempt at increasing efficiency and turnover due to financial distress.

Despite the generally low prevalence of loneliness in the sampled population there are important patterns appearing that generally support our hypotheses with a few exceptions.

## Relationship between loneliness, social contacts, and farm structures (hypotheses)

**Loneliness is inversely correlated with the frequency of social contacts (H1).** Results support hypothesis 1 and show an inverse correlation between loneliness and total social contacts (H1A), personal contacts (H1B), and professional contacts. Among personal contacts, those with colleagues and participation in social organizations have the strongest inverse association with loneliness. The frequency of loneliness feelings also correlates with greater *losses* in personal contacts with friends, family, and colleagues over the last two decades. In summary, the frequency of close social contacts—assumingly those with a certain quality/profoundness—and not the number of different social contacts, seems to have a direct bearing on loneliness.

Our finding that the number of different social contacts exhibits a positive correlation with loneliness may seem counterintuitive at first, since we would expect a higher diversity of social contacts to be correlated with less frequent feelings of loneliness. A possible reason is that the variable *Count* includes predominantly professional contacts and can be interpreted as a proxy for administrative complexity or farming complexity; however, without a better measure of such complexity, this explanation remains at the level of a hypothesis. Another surprising finding is the connection between loneliness and political contacts, which might signify that distressed farmers are more likely to be politically engaged, or perhaps that the political engagement leaves less room for other social interactions.

**Loneliness is correlated with higher workloads (H2).** Our findings support H2 that higher workloads correlate with more frequent feelings of loneliness. The PA analysis suggests that there are two pathways that link loneliness and workloads, namely a direct pathway (H2B), and an indirect pathway through the inverse association between workloads and personal contacts (H2A).

Other studies also report a relationship between time constraints, associated with higher workloads (Umstätter et al., 2022), and mental health issues among farmers. Based on structural equation modeling approaches, Logstein (2016) found a strong association between Norwegian farmers' load of mental health complains on one hand, and time concerns and economic concerns on the other, with no other farm characteristics predicting high symptoms of mental complaints. Notably, increases in farmers' workloads *additively* associated with higher levels of time concern *and* mental complaints, but workloads were associated with mental complaints *primarily indirectly* through higher time concerns (workload→time concern→mental complaint) (Logstein, 2016). Logstein's results are compelling in that they reveal that higher workloads *per se* are not necessarily associated with mental health issues; rather the concern that impinges on mental health is not having enough time to complete all the farm duties.

The relative effect of time management pressures on farmers' mental health differs between studies. Most studies highlight the importance of economic concerns over other types of stressors. Logstein (2016) shows that economic concerns are more strongly associated with mental health issues than time concerns. In a review by Yazd, Wheeler, and Zuo (2019), finances are the most cited (18%) farmer mental health risk factor in developing countries, whereas time pressure is among the least cited (4%). In a qualitative study on famers in the US Midwest region, five broad themes emerged as the main drivers of stress, namely finances, planning, weather, family concerns, and healthcare—with time management being just one of several challenges related to planning (Henning-Smith et al., 2022). However, other studies assign a higher weight to time management pressures. A quantitative study on US Midwest farmers found that time pressure was the second most-important stress domain following finances, and surpassing other concerns such as macroeconomic conditions, employee relationships, weather, hazardous work conditions, or social isolation (Rudolphi, Berg, and Parsaik 2020). In this work, we could not test the relationship between loneliness and financial factors, since we do not have data on the latter.

**Loneliness is inversely correlated with agricultural area (H3).** Our result that managers of larger farms feel lonely less frequently partly confirm H3. We had hypothesized that managers of larger farms feel less lonely *because they have more frequent social contacts* Junquera et al. (2022). However, PA results show that there is a significant inverse correlation between loneliness and agricultural area without the mediation of social contacts. A possible explanation is that a smaller agricultural area increases the need for off-farm income. As a consequence, farm managers may have even less time to maintain social contacts and may have fewer opportunities for social contacts, since they spend most of the working hours away from the farm and once home they spend their time completing farm-related tasks rather than socializing. A higher reliance on nonagricultural income may also negatively affect farmers' self image. However, this possible explanation remains at the level of a hypothesis which we did not test here given our lack of data on agricultural and non-agricultural income.

Our hypothesis that higher farming intensity is associated with less frequent feelings of loneliness is not confirmed. On one hand, managers of more intensive farms have more frequent social contacts, but on the other, they have also lost more personal contacts over the last two decades (Junquera et al. 2022), which could partly explain the lack of correlation between loneliness and farming intensity.

**Loneliness is inversely correlated with certifications or labels of local origin or organic agriculture (H4).**
This relationship is both direct (H4B), as well as indirect through more frequent personal and
professional contacts (H4A) for farms with the local origin label. Renting et al. (2003) refer to local origin
labels as a type of short food supply chain, which not only differentiates products from "anonymous"
commodities but also helps embed farming by establishing connections along the supply chain, for
example with local customers (e.g., through direct sales) and businesses such as local stores or dairies.
Presumably, such contacts may reduce social loneliness. Being part of a local origin label or organic
certification may also provide farmers with a sense of belonging and shared values (Cuéllar-Padilla and
Calle-Collado, 2011), and may thus reduce farmers' feelings of cultural loneliness as defined by (Wheeler
et al., 2023), although this question has received little attention in the literature so far.

**No correlation with physical isolation (H5).** Physical isolation, as measured by the number of neighbors
within a certain radius or the distance to the nearest neighbor, exhibits no correlation with loneliness,
which contradicts Hypothesis 5. This result could have two reasons. First, the population density in the
UBE is relatively high with 43 residents/km2 in 2021 (LUSTAT, 2023) compared to less than 5
persons/km2 in much of the US Midwest and West in 2020 (US Census Bureau, 2023). In densely
populated areas, the presence of neighbors may not be the limiting factor affecting loneliness: there
may be enough people in the vicinity, but the limiting factor may be finding the time to socially interact.
Second, the number of neighbors may influence social loneliness, but it does necessarily influence
emotional loneliness or cultural loneliness, and it has no bearing on the quality of social contacts.
Similarly, Buecker et al. (2021) report that the number of potential social contacts, i.e., population
density, does not predict loneliness in a nationally-representative German socio-economic panel.
Studies focusing on urban areas have even found a significant positive association between population
density and loneliness (Lai et al., 2021; MacDonald et al., 2020). The other measure of physical isolation
in our data is elevation zone. This variable exhibits a trend of higher loneliness at higher elevations, but
the relationship is not significant.

## Loneliness and structural change: are they related?

Our results suggest that structural change likely has a negative bearing on feelings of loneliness among
farmers as a result of increasing per-person workloads. Higher workloads can increase time
management pressures and associated stress, and leave less time for personal social exchanges. These
findings partly support the Goldschmid hypothesis, in the sense of a negative relationship between a
measure of farm size and social connectivity, although they differ from Goldschmid's in that the relevant measure is workload and not agricultural area. In fact, our finding that smaller agricultural area correlates with higher loneliness—controlling for workload—is worth exploring further and could point at a problem of financial viability below a certain farm size, and/or at decreasing social capital associated with smaller, less intensive and less financially viable operations.

## Our results in perspective

**Farmers vs. non-farmers.** A question of interest is whether the frequency of loneliness feelings among farmers differs significantly from that of the general population. Richard et al. (2017) examined self-assessed loneliness levels in the Swiss population based on responses from 20,007 participants in the cross-sectional Swiss Health Survey of 2012 and found that 64.1% of Swiss people are never lonely, and the rest (35.9%) is lonely sometimes (31.7%), quite often (2.7%), or very often (1.5%).  Unfortunately, Richard et al.'s four-point Likert scale (never, sometimes, quite often, very often) is not directly comparable to our six-point scale. Yet considering that Richard et al.'s category "sometimes" overlaps with our categories "rarely" and "sometimes", our results seem to align well with those of Richard et al. and suggest that loneliness among farmers in the sampled population does not differ strongly from the Swiss average. However, the different methodological approaches do not allow us to make a precise statement to this effect. Also, the study by Richard et al. does not compare loneliness between occupational groups or across the rural-urban spectrum, but rather focuses on lifestyle, age, and physical and mental health characteristics.

Among reviews comparing famers vs. non-farmers with regards to mental health (Yazd et al., 2019), burnout (Shaunessy 2022), and suicide in the US (Kennedy et al., 2021) and Switzerland (Steck et al., 2020), we did not find studies specifically focused on loneliness.  However, it is noteworthy that all of these studies report higher incidence of mental health issues among farmers than among non-farmers.

**Entlebuch vs. other regions.** Our results also need to be put in perspective with regards to the exceptionality of our selected study area. In the UBE, farming still constitutes an important share of the economy and plays a central role in local society and culture. This may also mean that farmers' social networks in the UBE are stronger than in other regions.  Our findings may thus not be representative of loneliness among farmers in other rural areas and may paint perhaps an overly positive picture.

Farmers in the UBE are also relatively "well-off" with regards to structural change compared to farmers in other parts of Switzerland and Europe. In the European Union, the number of farms declined by about 37% percent between 2005 and 2020, corresponding to the loss of 5.3 million farms, the vast majority (87%) of which were small farms under 5 ha. Yet the amount of agricultural land remained approximately steady, which means that the average mean size of agricultural holdings increased, from 11 ha in 2005 to 17.4 in 2020 (eurostat, 2023). In the same period, farm numbers declined in Switzerland by 22% while agricultural area grew by 26% (BfS 2023). In the UBE, farm numbers decreased by 18% from 1015 to 827, and agricultural area per farm increased by 18% from 14.8 ha to 17.6 ha (BfS, 2023b).

Structural change is thus slower in the UBE compared to the rest of Switzerland and other European regions. The explanation for this is complex, but one of the reasons is likely the tradition-bound culture in the UBE, with farming being a central element of the cultural identity. Whereas many farms are on the fringe of economic profitability, the drive to preserve the family farm is high. Such an attachment to the farming life and lifestyle may also come with more social engagement, and the many social organizations registered in the UBE testify to this. All of this may contribute to overall lower loneliness levels among farmers in this region compared to the rest of Switzerland—a question that could be explored in future studies.

## Limitations

As Wheeler et al. (2023) point out, loneliness has many dimensions that play into each other. Our collected data can only capture some of the dimensions of loneliness. This is reflected in the relatively low adjusted $R^2$ in the multiple linear regression analyses, which shows that our predictors only explain up to 10% of the variation in loneliness.  Similarly, the regression model in MacDonald et al. (2020) accounts for 2.8% of variation.

Our explanatory variables best explain some aspects of social loneliness, associated with social isolation, and to a lesser extent some aspects of emotional loneliness, associated with the quality of social contacts. With regards to emotional loneliness, we have assumed that contacts with family, friends, and colleagues are closer and more emotionally fulfilling than, say, contacts with the local butcher; however, we do not have data to corroborate this, as we did not ask this in the questionnaire. In other words, an unexplored aspect in our questionnaire is the quality of the contacts and the frequency of individually perceived close contacts and its change over the last decades. These aspects would need to be elucidated in further investigations. Our data also does not capture cultural loneliness. Yet these aspects may play a growingly important role, as farmers are increasingly portrayed as climate and environmental "perpetrators" in the media and public discourse , and as they become increasingly reliant on subsidies for their survival as well as more and more scarce due to the ongoing structural changes. There are also many other factors that affect loneliness beyond social contacts, including psychological aspects (Jackson-Smith and Gillespie, 2005) and stress factors such as financial concerns or fear of losing the farm.  These are key issues to be investigated further.

Finally, our data only capture loneliness at one point in time. It would be important to investigate the trends in loneliness and social relations, e.g., with a panel study.  This would help produce more robust findings about the relationship between structural change, social connections, and loneliness among farmers.

## CONCLUSIONS

The Swiss Center for Agricultural Research Agroscope recently observed that structural change in agriculture is a trend that is unavoidable, less acute than in neighboring European countries, and, thanks to public subsidies, remains within the bounds of what is deemed socially acceptable, namely a 2.5% annual reduction in farm numbers (Zorn, 2020). Our present and previous studies shed light on the fact that structural change in agriculture does not simply imply, as its name suggests, a physical transformation of farm structures. Structural change in agriculture, as we show, is closely connected to changes in social structures between farmers with implications at the level of the individual and the community. At the individual level, they entail a progressive reduction in the frequency of personal and local social contacts and may impact or compound farmers' feelings of loneliness. At the community level, this growing individualization can entail a loss of social capital in the form of local knowledge exchange, feelings of belonging, voluntary engagement, etc.

Our work highlights that when redesigning or evaluating agricultural policies, social externalities as described above must be taken into account, as they can have far reaching impacts in rural regions. Local authorities can foster social connectedness through interventions such as subsidizing organizations, organizing events and round-tables, or creating public spaces for social exchange. As we show, local food labels also seem to have multiple positive outcomes. However, these local interventions do not address some of the systemic roots of changing social relations. Our study highlights the social impacts of sustained structural change at the individual, community, and higher levels. Such impacts may be subtle in the short term, but important and cumulative in the longer term, and they should be considered in the design of policies affecting rural landscapes. Decision makers should be aware that public support for "livable" farming conditions not only sustains the economic integrity of the agricultural sector, but also the social fabric of rural regions.

Finally, the relatively "low" number of farm managers citing feeling lonely often or very often (three out of 104) should not be a reason for lowering the guard. Constituting almost 3% of the sampled population, these farmers could be at a higher risk for mental health issues, and such numbers might be higher outside the UBE in regions with less vibrant rural communities.